\documentclass[review]{elsarticle}

\usepackage[latin1]{inputenc}
\usepackage[T1]{fontenc} %\frenchspacing
\usepackage{amsmath,amssymb,url}
\usepackage{graphicx,float, pifont}
\usepackage{floatflt,subfigure,longtable}
\usepackage[active]{srcltx,rotating}
\usepackage{slashbox}
\usepackage{bm}
\usepackage{mathrsfs}
\usepackage{color}

\usepackage{lineno,hyperref}
\modulolinenumbers[5]

\journal{Journal of \LaTeX\ Templates}

\def\ep{\varepsilon}
\def\dfrac{\displaystyle\frac}
\def\ep{\varepsilon}

\begin{document}

\begin{frontmatter}

\title{Finite Element Analysis of Electromagnetic Waves in Two-Dimensional Transformed
Bianisotropic Media}

%\author{Yan Liu$^{a,^*}$, Boris Gralak$^b$, Sebastien Guenneau$^b$}
%\address{$^a$ School of Aerospace Science and Technology, Xidian University\\
%710126 Xi'an, China \\
%$^b$ Aix-Marseille Universit\'e, CNRS, Ecole Centrale Marseille, Institut Fresnel\\
%13013 Marseille, France \\}
%
%%\renewcommand{\thefootnote}{\fnsymbol{footnote}}
%\fntext[1]{Corresponding author.\\
%\,E-mail address: yanliu@xidian.edu.cn}

\author[mymainaddress]{Yan Liu\corref{mycorrespondingauthor}}
\cortext[mycorrespondingauthor]{Corresponding author}
\ead{yanliu@xidian.edu.cn}

\author[mysecondaryaddress]{Boris Gralak}
\author[mysecondaryaddress]{Sebastien Guenneau}

\address[mymainaddress]{School of Aerospace Science and Technology, Xidian University,
Xi'an 710071, China}
\address[mysecondaryaddress]{Aix-Marseille Universit\'e, CNRS, Ecole Centrale Marseille, Institut Fresnel\\
13013 Marseille, France}

\begin{abstract}
We analyse wave propagation in two-dimensional bianisotropic media with the Finite Element Method (FEM). We start from the Maxwell-Tellegen's equations in bianisotropic media, and derive some system of coupled Partial Difference Equations (PDEs) for longitudinal electric and magnetic field components. Perfectly Matched Layers (PMLs) are discussed to model such unbounded media. We implement
these PDEs and PMLs in a finite element software. We apply transformation optics in order to design some bianisotropic media with interesting functionalities, such as cloaks, concentrators and rotators. We propose a design of metamaterial with concentric layers
made of homogeneous media with isotropic permittivity, permeability and magneto-electric parameters that mimic
the required effective anisotropic tensors of a bianisotropic cloak in the long wavelength limit (homogenization approach).
Our numerical results show that well-known metamaterials can be transposed to bianisotropic media.
\end{abstract}


\end{frontmatter}

%\linenumbers

\section{Introduction}
In the last decade, metamaterials have attracted much attention due to their extraordinary properties, such as negative refraction, ultra refraction, anomalous dispersion and so on. Metamaterials are artificial materials engineered to gain desired properties that cannot be found in nature and usually consist of periodically arranged materials, which affect electromagnetic waves in an unconventional manner. More precisely, they exhibit new and unusual electromagnetic properties at the macro-scale, due to their structural features smaller than the operational wavelength of the electromagnetic wave. Metamaterials include the negative index materials (NIMs), single negative metamaterials, bi-isotropic and bianisotropic media, chiral media and so on.

As an application, the physicist John Pendry proposed that NIMs enable a perfect lens \cite{Pendry00}, which allows a sub-wavelength imaging. On the other hand, J. Pendry et al in 2006 \cite{Pendry06,Leonhardt06} proposed that a geometric transformation of space could distort light trajectories around a bounded region, which is thus made invisible. This powerful mathematical technique is called transformation optics, and it presents great potential thanks to the fabrication of metamaterials with specially designed properties \cite{Rahm08,Rahm2008,Chen07}. Transformation optics has been successfully applied to design functionalities, such as invisibility cloaks, rotators, concentrators etc.

In NIMs, or single negative metamaterials (e.g. composites with $\ep<0$ or $\mu<0$), we have supposed that the metamaterials have independent electric and magnetic responses described by $\ep$ and $\mu$. However, magnetoelectric coupling does occur in many other electromagnetic metamaterials, wherein the electric and magnetic fields are induced reciprocally. Anisotropic (resp. isotropic) media with such kind of property are referred as bianisotropic (resp. bi-isotropic) \cite{Marques02,Rill09}.

In \cite{yanl13,yanl14}, we have theoretically proved that the Pendry-Ramakrishna lens theorem is applicable to complementary bianisotropic media; on the other hand, as a powerful mathematical technique - transformation optics allows for a transformation of space between two different coordinate systems, while the Maxwell-Tellegen's equations for bianisotropic media are also proved to be form invariant under space transformation. Although it is easy to understand the principle used to design novel devices such as cloaks, concentrators, rotators and so on, a numerical model is required to illustrate their electromagnetic response. Hence, in this paper, we would like to discuss in details the implementation of transformed bianisotropic media in a finite element model, and further
show numerical results for an invisibility cloak, concentrator and rotator, the former being even
achieved with a simplified multilayered design.

We know that, Partial Differential Equations (PDEs) are generally used to model many physical phenomena such as fluid dynamics and electromagnetism, etc. In complex media, solutions to the governing equations can be difficult to derive in closed-form by traditional analytical routes - Fourier or Laplace transform methods, power series expansion and so on, hence one often has to resort to
numerical approximation of the solutions. As a particular class of numerical techniques, FEM is efficient to solve problems in
heterogeneous anisotropic media \cite{Courant43,Zlamal68}, and it has been widely applied in engineering design and analysis
in mechanics starting from the late seventies, while researchers in photonics started to use it in the nineties.
Notably, the COMSOL Multiphysics package is a finite element analysis software much used nowadays to solve various
physics and engineering applications in the metamaterials' community, as it allows entering coupled systems of PDEs,
such as in opto-mechanics, thermal elasticity etc. Here, we make use of it to set up a coupled PDE system for a bianisotropic
structure. To do this, we start from the Maxwell-Tellegen's equations in bianisotropic media and derive two coupled PDEs with the longitudinal electric and magnetic fields as the unknowns; then we discuss open boundary conditions known in the photonics literature as perfectly matched layers (PMLs). To illustrate the usefulness of our numerical algorithm, we apply transformation optics to bianisotropic
media in order to design some functional devices, i.e. invisibility cloak, concentrator and rotator. Their electromagnetic
(EM) properties studied thanks to our PDEs model implemented in COMSOL Multiphysics prove that these
well-known devices work equally well in bianisotropic media.

% for bianisotropic media are deduced starting from the Maxwell-Tellegen's equations; then the boundary conditions is discussed, in particularly, an open boundary condition is mimicked by proper perfect matched layers (PMLs); finally, as a numerical illustration, the optics transformation in bianisotropic media is introduced to design a series of functional devices, i.e. invisibility cloak, concentrator and rotator, and the EM properties of these structures are analyzed.
%--------------------------------------------------------------------------------------------------------------------------------
\section{A coupled PDE sytem for bianisotropic media}
We first recall the time-harmonic Maxwell's equations (assuming a time dependence $e^{-i\omega t}$
with $t$ the time variable and $\omega$ the angular wave frequency)
\begin{equation}
\begin{array}{ll}
\nabla \times {\bm E} &= i \omega {\bm B}\,,\\
\nabla \times {\bm H} &= -i \omega {\bm D}\,.\\
\end{array}
\label{f3}
\end{equation}
where ${\bm E}$ and ${\bm H}$ are the electric and magnetic field respectively, while ${\bm D}$ and ${\bm B}$ are the electric displacement and magnetic flux density.

Let us assume the constitutive relations in a bianisotropic medium described by:
\begin{equation}
\begin{array}{ll}
{\bm D} &= \ep {\bm E}+i \xi {\bm H}\,,\\
{\bm B} &= -i \zeta{\bm E}+\mu {\bm H}\,.\\
\end{array}
\label{f4}
\end{equation}
where $\ep$ is the permittivity, $\mu$ is the permeability, $\xi$ and $\zeta$ are the magnetoelectric coupling parameters.
All these material parameters can be treated as rank-2 tensors. We now substitute (\ref{f4}) into (\ref{f3}) and obtain the Maxwell-Tellegen's equations
\begin{equation}
\begin{array}{lll}
\nabla \times {\bm E} &= \omega \zeta {\bm E}+i \omega \mu {\bm H}\,,\\
\nabla \times {\bm H} &= -i \omega \varepsilon {\bm E}+\omega \xi {\bm H}\,.\\
\end{array}
\label{f5}
\end{equation}

Let us now consider an orthogonal coordinate system $(x_1,x_2,x_3)$, and assume that the material parameters and the electromagnetic field are invariant in one direction, say, the $x_3$ direction, that is, $\varepsilon$, $\mu$, $\xi$, ${\bm E}$, ${\bm H}$ are independent
of $x_3$.

Let us define ${\bm E} = \left(E_1,E_2,E_3\right)^{\rm T}$, ${\bm H} = \left(H_1,H_2,H_3\right)^{\rm T}$, and further assume following the terminology used in the book by F. Zolla et al. \cite{Zolla12}, that $\varepsilon$, $\mu$, $\xi$ and $\zeta$ are $z$-anisotropic tensors i.e. such that
\begin{equation}
\begin{array}{ll}
\varepsilon =\left[\begin{array}{ccc}
\varepsilon_{11} & \varepsilon_{12} & 0\\
\varepsilon_{21} & \varepsilon_{22} & 0\\
0 & 0 & \varepsilon_{33}
\end{array}\right] , &
\mu =\left[\begin{array}{ccc}
\mu_{11} & \mu_{12} & 0\\
\mu_{21} & \mu_{22} & 0\\
0 & 0 & \mu_{33}
\end{array}\right] , \\[1cm]
\xi =\left[\begin{array}{ccc}
\xi_{11} & \xi_{12} & 0\\
\xi_{21} & \xi_{22} & 0\\
0 & 0 & \xi_{33}
\end{array}\right]\,,&
\zeta =\left[\begin{array}{ccc}
\zeta_{11} & \zeta_{12} & 0\\
\zeta_{21} & \zeta_{22} & 0\\
0 & 0 & \zeta_{33}
\end{array}\right]\,.
\end{array}
\label{f10a}
\end{equation}

We then apply (\ref{f10a}) to (\ref{f5}), and have
\begin{equation}
\partial_{x_2} E_3 = \omega \zeta_{11} E_1+\omega \zeta_{12} E_2+i\omega \mu_{11} H_1+i\omega \mu_{12} H_2\,,
\label{f11-1}
\end{equation}
\vspace{-0.5cm}
\begin{equation}
-\partial_{x_1} E_3 = \omega \zeta_{21} E_1+\omega \zeta_{22} E_2+i\omega \mu_{21} H_1+i\omega \mu_{22} H_2\,,
\label{f11-2}
\end{equation}
\vspace{-0.5cm}
\begin{equation}
\partial_{x_1} E_2-\partial_{x_2} E_1 = \omega \zeta_{33} E_3+i\omega \mu_{33} H_3\,,
\label{f11-3}
\end{equation}
\vspace{-0.5cm}
\begin{equation}
\partial_{x_2} H_3 = -i\omega \varepsilon_{11} E_1-i\omega \varepsilon_{12} E_2+\omega \xi_{11} H_1+\omega \xi_{12} H_2\,,
\label{f12-1a}
\end{equation}
\vspace{-0.5cm}
\begin{equation}
-\partial_{x_1} H_3 = -i\omega \varepsilon_{21} E_1-i\omega \varepsilon_{22} E_2+\omega \xi_{21} H_1+\omega \xi_{22} H_2\,,
\label{f12-2a}
\end{equation}
\vspace{-0.5cm}
\begin{equation}
\partial_{x_1} H_2-\partial_{x_2} H_1 = -i\omega \varepsilon_{33} E_3+\omega \xi_{33} H_3\,.
\label{f12-3a}
\end{equation}
We rewrite (\ref{f11-1}) and (\ref{f11-2}) in a matrix form
\begin{equation}
\left[ \begin{array}{cc}
\partial_{x_1} E_3\\[2mm]
\partial_{x_2} E_3
\end{array}\right]
=\omega \left[\begin{array}{cc}
-\zeta_{22} & \zeta_{21}\\
\zeta_{12} & -\zeta_{11}
\end{array}\right]
\left[\begin{array}{cc}
E_2 \\
-E_1
\end{array}\right]+i\omega\left[\begin{array}{cc}
-\mu_{22} & \mu_{21}\\
\mu_{12} & -\mu_{11}
\end{array}\right]
\left[\begin{array}{cc}
H_2 \\
-H_1
\end{array}\right]\,.
\label{f13-1}
\end{equation}
Similarly, (\ref{f12-1a}) and (\ref{f12-2a}) become
\begin{equation}
\left[ \begin{array}{cc}
\partial_{x_1} H_3\\[2mm]
\partial_{x_2} H_3
\end{array}\right]
= i\omega \left[\begin{array}{cc}
\varepsilon_{22} & -\varepsilon_{21}\\
-\varepsilon_{12} & \varepsilon_{11}
\end{array}\right]
\left[\begin{array}{cc}
E_2 \\
-E_1
\end{array}\right]+\omega\left[\begin{array}{cc}
-\xi_{22} & \xi_{21}\\
\xi_{12} & -\xi_{11}
\end{array}\right]
\left[\begin{array}{cc}
H_2 \\
-H_1
\end{array}\right]\,.
\label{f13-2}
\end{equation}
If we define
\begin{equation}
\begin{array}{ll}
\varepsilon_T=\left[\begin{array}{cc}
\varepsilon_{22} & -\varepsilon_{21}\\
-\varepsilon_{12} & \varepsilon_{11}
\end{array}\right] ,&
\mu_T=\left[\begin{array}{cc}
-\mu_{22} & \mu_{21}\\
\mu_{12} & -\mu_{11}
\end{array}\right] ,\\[0.5cm]
\xi_T=\left[\begin{array}{cc}
-\xi_{22} & \xi_{21}\\
\xi_{12} & -\xi_{11}
\end{array}\right],&
\zeta_T=\left[\begin{array}{cc}
-\zeta_{22} & \zeta_{21}\\
\zeta_{12} & -\zeta_{11}
\end{array}\right],
\end{array}
\label{f14}
\end{equation}
and
\begin{equation}
\underline{A}=\left[ \begin{array}{cc}
\partial_{x_1} \\
\partial_{x_2}
\end{array}\right],\quad
{\underline{E}}=
\left[\begin{array}{cc}
E_2 \\
-E_1
\end{array}\right],\quad
{\underline{H}}=
\left[\begin{array}{cc}
H_2 \\
-H_1
\end{array}\right]\,.
\label{f15}
\end{equation}
then (\ref{f13-1}) and (\ref{f13-2}) turn out to be
\begin{equation}
\underline{A}E_3 =  \omega \zeta_T \underline{E}+i \omega \mu_T \underline{H}\,,
\label{f16-1}
\end{equation}
\vspace{-0.5cm}
\begin{equation}
\underline{A}H_3 =  i\omega \varepsilon_T \underline{E}+ \omega \xi_T \underline{H}\,.
\label{f16-2}
\end{equation}
Furthermore, from (\ref{f16-1}), we have
\begin{equation}
\underline{E}= {\omega}^{-1} \zeta^{-1}_T\underline{A}E_3-i \zeta^{-1}_T \mu_T \underline{H}\,.
\label{f17}
\end{equation}
Substituting it into (\ref{f16-2}), we obtain
\begin{equation}
\begin{array}{ll}
\underline{A}H_3 &=i\omega \varepsilon_T ({\omega}^{-1} {\zeta^{-1}_T} \underline{A}E_3
-i \zeta^{-1}_T \mu_T \underline{H})+ \omega \xi_T \underline{H}\\[0.2cm]
&=i\varepsilon_T\; \zeta^{-1}_T \underline{A}E_3
+ \omega \varepsilon_T\; \zeta^{-1}_T \;\mu_T \underline{H}
+ \omega \xi_T \underline{H}\,.
\end{array}
\label{f18}
\end{equation}
which can be recast as
\begin{equation}
\underline{H} = B^{-1}\left( \underline{A}H_3 - i \varepsilon_T\;
\zeta^{-1}_T\underline{A}E_3 \right)\,,
\label{f19}
\end{equation}
with
\begin{equation}
B= \omega \xi_T + \omega \varepsilon_T\;
\zeta^{-1}_T \mu_T\,.
\label{f19-1}
\end{equation}
Let us now plug (\ref{f19}) into (\ref{f12-3a})
\begin{equation}
\begin{array}{ll}
\left[\begin{array}{cc}
\partial_{x_1} & \partial_{x_2}
\end{array}\right] \underline{H} &=
\left[\begin{array}{cc}
\partial_{x_1} & \partial_{x_2}
\end{array}\right]
\left(B^{-1} \underline{A}H_3 - iB^{-1}\;\varepsilon_T
\zeta^{-1}_T \underline{A}E_3\right) \\[0.2cm]
&=\left[\begin{array}{cc}
\partial_{x_1} & \partial_{x_2}
\end{array}\right]  \left\{-\dfrac{i}{\omega}(\mu_T+\zeta_T
\varepsilon^{-1}_T \xi_T )^{-1}\right\} \underline{A}E_3  \\[0.2cm]
&+\left[\begin{array}{cc}
\partial_{x_1} & \partial_{x_2}
\end{array}\right] \left\{\dfrac{1}{\omega}(\xi_T + \varepsilon_T
\zeta^{-1}_T\;\mu_T)^{-1}\right\} \underline{A}H_3 \\[0.2cm]
&=-i\omega \varepsilon_{33}E_3+\omega \xi_{33}H_3 \,.
\label{f19-2}
\end{array}
\end{equation}
In the same way, from (\ref{f16-2}), we have
\begin{equation}
\underline{H}= {\omega^{-1}} \xi^{-1}_T \underline{A}H_3
- i \xi^{-1}_T \varepsilon_T \underline{E}\,,
\label{f20}
\end{equation}
which is then applied to (\ref{f16-1})
\begin{equation}
\begin{array}{ll}
\underline{A}E_3 &= \omega \zeta_T\underline{E} + i\omega \mu_T {({\omega^{-1}}
\xi^{-1}_T \underline{A}H_3 - i \xi^{-1}_T \varepsilon_T \underline{E})} \\[0.2cm]
&=\omega \zeta_T \underline{E}+i \mu_T \;\xi^{-1}_T \underline{A}H_3+
   \omega \mu_T\; \xi^{-1}_T \varepsilon_T \underline{E}\,,
\label{f21}
\end{array}
\end{equation}
where
\begin{equation}
\underline{E} = C^{-1} \underline{A}E_3 - i C^{-1} \mu_T\;
 \xi^{-1}_T \underline{A}H_3\,,
\label{f22}
\end{equation}
with
\begin{equation}
C = \omega \zeta_T + \omega \mu_T\;
\xi^{-1}_T \varepsilon_T\,.
\label{f22-1}
\end{equation}
We plug (\ref{f22}) into (\ref{f11-3})
\begin{equation}
\begin{array}{ll}
[\begin{array}{cc}
\partial_{x_1} & \partial_{x_2}
\end{array}] {\underline{E}} &=
[\begin{array}{cc}
\partial_{x_1} & \partial_{x_2}
\end{array}]
\left(C^{-1} \underline{A}E_3-iC^{-1} \mu_T\;
 \xi^{-1}_T \underline{A}H_3\right)  \\[0.2cm]
&=[\begin{array}{cc}
\partial_{x_1} & \partial_{x_2}
\end{array}] \left\{\dfrac{1}{\omega}(\zeta_T + \mu_T
\xi^{-1}_T\;\varepsilon_T)^{-1}\right\} \underline{A}E_3 \\[0.2cm]
&-[\begin{array}{cc}
\partial_{x_1} & \partial_{x_2}
\end{array}]  \left\{\dfrac{i}{\omega}(\varepsilon_T+\xi_T
\mu^{-1}_T\;\zeta_T)^{-1}\right\} \underline{A}H_3 \\[0.2cm]
&=\omega \zeta_{33}E_3+i\omega \mu_{33}H_3\,.
\end{array}
\label{f23}
\end{equation}
Finally, we rewrite (\ref{f19-2}) and (\ref{f23}) by multiplying by $\omega$ on both sides of the equations, and
we obtain
\begin{equation}
\begin{array}{l}
\left[\begin{array}{cc}
\partial_{x_1} & \partial_{x_2}
\end{array}\right]  \left\{-i(\mu_T+\zeta_T\;
\varepsilon^{-1}_T\;\xi_T)^{-1}\right\}
\left[ \begin{array}{cc}
\partial_{x_1} & \partial_{x_2}
\end{array}\right]^{\rm T}E_3 \\[0.2cm]
+\left[\begin{array}{cc}
\partial_{x_1} & \partial_{x_2}
\end{array}\right] \left\{(\xi_T + \varepsilon_T\;
\zeta^{-1}_T\; \mu_T)^{-1}\right\}
\left[ \begin{array}{cc}
\partial_{x_1} & \partial_{x_2}
\end{array}\right]^{\rm T}H_3=-i\omega^2 \varepsilon_{33}E_3+\omega^2 \xi_{33}H_3 \,,\\[0.2cm]
\left[\begin{array}{cc}
\partial_{x_1} & \partial_{x_2}
\end{array}\right]
\left\{(\zeta_T + \mu_T\; \xi^{-1}_T \;
\varepsilon_T)^{-1}\right\}
\left[ \begin{array}{cc}
\partial_{x_1} & \partial_{x_2}
\end{array}\right]^{\rm T}E_3 \\[0.2cm]
+\left[\begin{array}{cc}
\partial_{x_1} & \partial_{x_2}
\end{array}\right]
\left\{-i(\varepsilon_T+\xi_T\; \mu^{-1}_T\;
\zeta_T)^{-1}\right\}
\left[ \begin{array}{cc}
\partial_{x_1} & \partial_{x_2}
\end{array}\right]^{\rm T}H_3=\omega^2 \zeta_{33}E_3+i\omega^2 \mu_{33}H_3 \,.
\end{array}
\label{f24}
\end{equation}
which are two coupled equations with the longitudinal fields $E_3$ and $H_3$ as unknowns.

\section{Implementation of the PDE system in Comsol Multiphysics}
The standard form of PDEs in COMSOL Multiphysics is described by
the following equation reminiscent of a governing equation
in elasticity theory:
\begin{equation}
\nabla \cdot (-c \nabla {\bf u}- \alpha {\bf u} + \gamma)+ a{\bf u}+ \beta \cdot \nabla {\bf u} = f \,.
\label{PDE}
\end{equation}
where $c$ is a rank-4 tensor,
$\alpha$ is a rank $2$-tensor, $\gamma$ and $\beta$ are
vectors and $f$ is a forcing term.

We rewrite the formulae in (\ref{f24}) as
\begin{equation}
\begin{array}{l}
\left[\begin{array}{cc}
\partial_{x_1} & \partial_{x_2}
\end{array}\right]  \left\{(\mu_T+\zeta_T \;
\varepsilon^{-1}_T \;\xi_T)^{-1}\right\}
\left[ \begin{array}{cc}
\partial_{x_1} & \partial_{x_2}
\end{array}\right]^{\rm T}E_3 \\[0.2cm]
+\left[\begin{array}{cc}
\partial_{x_1} & \partial_{x_2}
\end{array}\right] \left\{i(\xi_T + \varepsilon_T\;
\zeta^{-1}_T\; \mu_T)^{-1}\right\}
\left[ \begin{array}{cc}
\partial_{x_1} & \partial_{x_2}
\end{array}\right]^{\rm T}H_3 =\omega^2 \varepsilon_{33}E_3+i\omega^2 \xi_{33}H_3 \,,\\[0.2cm]
\left[\begin{array}{cc}
\partial_{x_1} & \partial_{x_2}
\end{array}\right] \left\{i(\zeta_T + \mu_T\;
\xi^{-1}_T \;\varepsilon_T)^{-1}\right\}
\left[ \begin{array}{cc}
\partial_{x_1} & \partial_{x_2}
\end{array}\right]^{\rm T} E_3 \\[0.2cm]
+\left[\begin{array}{cc}
\partial_{x_1} & \partial_{x_2}
\end{array}\right] \left\{(\varepsilon_T+\xi_T\;
\mu^{-1}_T\; \zeta_T)^{-1}\right\}
\left[ \begin{array}{cc}
\partial_{x_1} & \partial_{x_2}
\end{array}\right]^{\rm T} H_3 =i\omega^2 \zeta_{33}E_3 -\omega^2 \mu_{33}H_3\,.
\end{array}
\label{f25}
\end{equation}
Comparing with (\ref{PDE}), if we consider the coupled longitudinal electric and magnetic fields, the unknown ${\bf u}$ in (\ref{PDE}) is
\begin{equation}
  {\bf u}= \left[\begin{array}{cc}
  E_3 \\
  H_3
  \end{array}
  \right]\,.
\label{u}
\end{equation}
and the coefficients are
\begin{equation}
c=\left[\begin{array}{cc}
c_{11} & c_{12}\\
c_{21} & c_{22}
\end{array}\right], \quad
a=\left[\begin{array}{cc}
\omega^2 \varepsilon_{33} & i \omega^2 \xi_{33}\\
i \omega^2 \zeta_{33} & -\omega^2 \mu_{33}
\end{array}\right]\,.
\label{coeff}
\end{equation}
with entries
\begin{align}
c_{11}= (\mu_T+ \zeta_T\; \varepsilon^{-1}_T \;\xi_T)^{-1} \,,\quad
c_{12}= i(\xi_T +\varepsilon_T\; \zeta^{-1}_T \; \mu_T )^{-1} \,,\notag \\
c_{21}= i(\zeta_T + \mu_T \; \xi^{-1}_T \; \varepsilon_T)^{-1} \,,\quad
c_{22}= (\varepsilon_T + \xi_T\; \mu^{-1}_T\; \zeta_T)^{-1}\,.
\label{c}
\end{align}
Simply stated, when the tensors of $\varepsilon, \mu, \xi, \zeta$ in (\ref{f14}) are
represented by diagonal matrices in a Cartesian coordinate system
\begin{equation}
  v={\rm diag} (\nu_{11}, \nu_{22}, \nu_{33}),\quad v=\ep,\mu,\xi,\zeta\,.
\end{equation}
then the entries in (\ref{c}) become
\begin{equation}
\begin{array}{c}
c_{11}
= \left[\begin{array}{cc}
\dfrac{-\varepsilon_{22}}{\mu_{22} \varepsilon_{22}-\xi_{22}\zeta_{22}} & 0 \\
0 & \dfrac{-\varepsilon_{11}}{\mu_{11} \varepsilon_{11}-\xi_{11}\zeta_{11}}
\end{array}\right] ,  \\
c_{12}
= \left[\begin{array}{cc}
\dfrac{i\zeta_{22}}{\mu_{22} \varepsilon_{22}-\xi_{22}\zeta_{22}} & 0 \\
0 & \dfrac{i\zeta_{11}}{\mu_{11} \varepsilon_{11}-\xi_{11}\zeta_{11}}
\end{array}\right]  \,,\\
c_{21}
=\left[\begin{array}{cc}
\dfrac{i\xi_{22}}{\mu_{22} \varepsilon_{22}-\xi_{22}\zeta_{22}} & 0 \\
0 & \dfrac{i\xi_{11}}{\mu_{11} \varepsilon_{11}-\xi_{11}\zeta_{11}}
\end{array}\right] , \\
c_{22}
= \left[\begin{array}{cc}
\dfrac{\mu_{22}}{\mu_{22} \varepsilon_{22}-\xi_{22}\zeta_{22}} & 0 \\
0 & \dfrac{\mu_{11}}{\mu_{11} \varepsilon_{11}-\xi_{11}\zeta_{11}}
\end{array}\right]\,.
\end{array}
\label{cexact}
\end{equation}
Note that each component of the coefficients $c_{ij}\, (i,j=1,2)$ is a fraction with a denominator $(\ep_{kk}\mu_{kk}-\xi_{kk}\zeta_{kk})\,(k=1,2)$. In order to avoid any infinite entries of $c$ while solving the PDEs, it should satisfy
\begin{equation}
  \ep_{kk}\mu_{kk}-\xi_{kk}\zeta_{kk}\neq 0,\,\, (k=1,2)
  \label{condt}
\end{equation}
{\bf Remark}: This condition is essential and generally is met since the magnetoelectric coupling parameters for bianisotropic media are usually small.
%%------------------------------------------------------------------------------------------------------------------------------------------------
%%-----------------------------------------------------------------------------------------------------------------------------------------------
\section{Problem setup with open boundary conditions}
In the last section, we have discussed the general PDEs in bianisotropic media, which allows us to introduce the PDEs model in COMSOL Multiphysics; however, in order to mimic the scattering and propagation properties of bianisotropic media in an open space, we need to consider the boundary conditions. We usually introduce a transformation of an infinite domain into a finite one, which should enforce that the wavelength is contracted to infinitely small values as it approaches the outer boundary of the transformed domain within a perfectly matched layer (PML). At this outer boundary, Dirichlet data could be set, thereby enforcing a vanishing field \cite{Berenger94,Teixeira98}.

The PML is shown to be equivalent to an analytic continuation of Maxwell's equations to complex variables's spatial domain \cite{Teixeira97}. The form of the electromagnetic field inside a PML region can be obtained from the solutions in real space from a mapping to a complex space through this analytic continuation. Physically speaking, an equivalent artificial medium is achieved from this map in the PML region that is described by complex, anisotropic and inhomogeneous parameters, even if the original ones are real, isotropic and homogeneous. The analytic continuation means to alter the eigenfunctions of Maxwell's equations in such a way that the propagating modes are mapped continuously to exponentially decaying modes, which allows for reflectionless absorption of electromagnetic waves. In other words, an absorbing medium dissipating the outgoing wave can be achieved through proper complex map: an artificial medium situated in a region can be added to a finite medium in order to mimic an open space within which the field is damped to a negligible value \cite{Agha08}. Importantly, the impedance of the equivalent medium is the same as that of the initial medium since all the parameters undergo the same transform, what ensures non-reflecting features on the interface between the medium and the PMLs.

F. L. Teixeira et.al \cite{Teixeira97,Teixeira98} derive the Maxwellian PML's for arbitrary bianisotropic and dispersive media. In the Cartesian coordinate system, if an analytic continuation is defined by the transformation function
\begin{equation}
  {\tilde u} = \int_{0}^{u} s_u(u')du'\,.
\end{equation}
with $s_u(u')$ the complex stretching variables \cite{Chew94} and $u$ stands for $x$, $y$, $z$. In the complex space, the nabla operator in Maxwell's equations changes as
\begin{equation}
\begin{array}{ll}
  \nabla \rightarrow {\tilde \nabla} &= {\hat x}\dfrac{\partial}{\partial {\tilde x}}+{\hat y}\dfrac{\partial}{\partial {\tilde y}}
  +{\hat z}\dfrac{\partial}{\partial {\tilde z}} \\[3mm]
  &={\hat x}\dfrac{1}{s_x}\dfrac{\partial}{\partial x}+{\hat y}\dfrac{1}{s_y}\dfrac{\partial}{\partial y}
  +{\hat z}\dfrac{1}{s_z}\dfrac{\partial}{\partial z}\,,
\end{array}
\end{equation}
it reads as
\begin{equation}
  {\tilde \nabla} = \overset{=}{S} \cdot \nabla\,,
\end{equation}
with
\begin{equation}
  \overset{=}{S}={\hat x}{\hat x}(\dfrac{1}{s_x})+{\hat y}{\hat y}(\dfrac{1}{s_y})+{\hat z}{\hat z}(\dfrac{1}{s_z})\,.
\end{equation}
since $s_u(u)$ and $\partial/\partial u'$ commute for $u\neq u'$, $\overset{=}{S}$ is a diagonal tensor, and $\det (\overset{=}{S})= (s_x s_y s_z)^{-1}$.

Based on this, we then can expand the Maxwell's equations in the new complex space, and derive that the PMLs with bianisotropic constitutive parameters are
\begin{equation}
  v_{\rm PML}=(\det \overset{=}{S})^{-1} [\overset{=}{S} \cdot v \cdot\overset{=}{S}],\quad v=\ep,\mu,\xi,\zeta\,.
\end{equation}
\begin{figure}[!htb]
    \centering
    \includegraphics[scale=0.7]{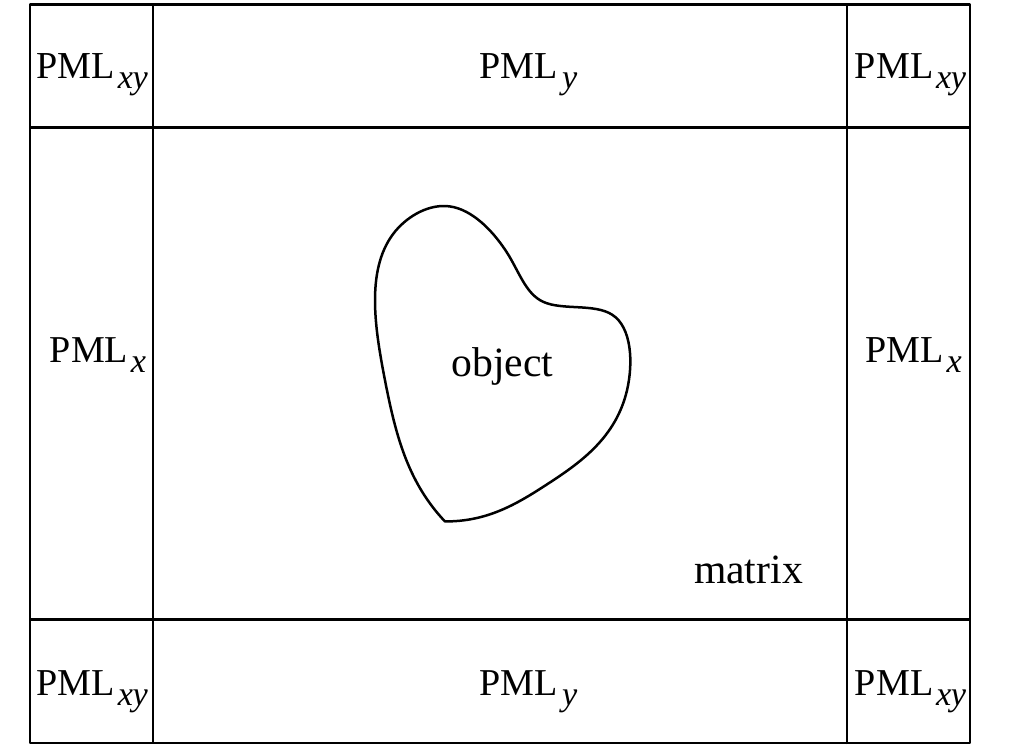}
    \caption{Computational domain enclosed by perfectly matched layers (PMLs) with functionality in the $x$, $y$ and $x$-$y$ direction \cite{Rahm08}.}
   \label{figpml}
\end{figure}

Take a full wave simulation for two dimensional structures in COMSOL Multiphysics as an illustration, Fig. \ref{figpml} shows the computational domain terminated by PMLs in Cartesian coordinates. The PMLs are designed to decrease the distribution of the waves along the $x$, $y$ and $x$-$y$ directions, respectively. Specifically, assuming the parameter of the matrix $v={\rm diag}(v_{xx},v_{yy},v_{zz})$ ($v=\ep,\mu,\xi,\zeta$), then we choose $\overset{=}{S}$
\begin{equation}
\begin{array}{llll}
  {\rm PML}_{x}: & s_x=a-b*i, &\, s_y=1, & \,s_z=1\,, \\
  {\rm PML}_{y}: & s_x=1, &\, s_y=a-b*i, & \,s_z=1\,, \\
  {\rm PML}_{xy}: & s_x=a-b*i, & \,s_y=a-b*i, & \,s_z=1\,.
\end{array}
\label{pml}
\end{equation}
with $a,b\in \mathbb{Z}^{+}$. Then the transformed parameters in the PMLs are
\begin{equation}
v'={\rm diag}(v_{xx} L_{xx},v_{yy}L_{yy},v_{zz}L_{zz})\,,\,\,\,v=\ep,\mu,\xi,\zeta\,.
\end{equation}
where
\begin{equation}
L_{xx}=\dfrac{s_y s_z}{s_x},\quad L_{yy}=\dfrac{s_x s_z}{s_y}, \quad L_{zz}=\dfrac{s_x s_y}{s_z}\,.
\end{equation}
can be derived by introducing the definition of $s_x$, $s_y$ and $s_z$ in (\ref{pml}) along different directions. Similar concept of PMLs can be extended to the anisotropic/bianisotropic matrix and other coordinate system \cite{Teixeira97,Teixeira1998}.
%%-----------------------------------------------------------------------------------------------------------------------------------------
%%------------------------------------------------------------------------------------------------------------------------------------------
\section{FEM implementation}
To check both our PDEs model and PMLs for bianisotropic media, some conceptual devices designed from transformation optics are numerically studied in the following subsections, these structures are all achieved by bianisotropic media.

\subsection{Invisibility cloak}
Electromagnetic (EM) metamaterials such as invisibility cloaks can be designed through the blowup of a point \cite{Leonhardt06,Pendry06}, using transformation optics which is a versatile mathematical tool enabling a deeper analytical insight into the scattering properties of EM fields in metamaterials. In this subsection, a cloak consisting of a bianisotropic medium is investigated, and the geometric transformation proposed by J. Pendry \cite{Pendry06} is proved to be applicable to design such kind of cloak.

\begin{figure}[!htb]
    \centerline{\includegraphics[width=0.95\textwidth]{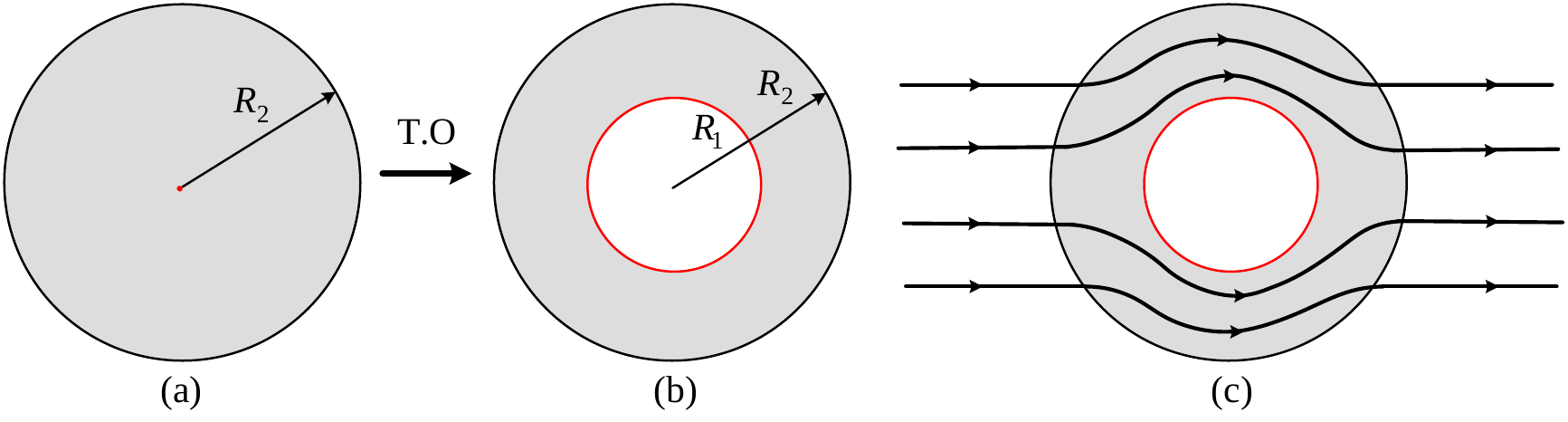}}
    \caption{Geometric transformation for invisibility cloak: (a) Disk (grey) with radius $r=R_2$ and the original point (red) is at $r=0$; (b) The original point is mapped to a disk with $r'=R_1$, i.e the disk with $r=R_2$ is compressed to an annulus $R_1 < r' \leq R_2$, the mapping function is defined by (\ref{gtcloak}); (c) Ray trajectories in the cloak. }
    \label{figcloakto}
\end{figure}

We start with a free space with permittivity $\ep=\ep_0$, permeability $\mu=\mu_0$ and magnetoelectric coupling parameters $\xi=\xi_0$,
$\zeta=\zeta_0$ in the Cartesian coordinate system, and first map it onto polar coordinates $(r,\theta,z)$ as defined by
\begin{equation}
  x=r\cos{\theta},\,\, y=r\sin{\theta},\,\, z=z.
  \label{gt1}
\end{equation}
a disk with $R_2$ results as shown in Fig. \ref{figcloakto}(a). Let us then introduce a geometric transformation defined by (\ref{gtcloak}) which maps the field within this disk onto an annulus $R_1< r \leq R_2$, i.e. the original point (red) $r=0$ in the original disk (Fig. \ref{figcloakto}(a)) is blowup to a disk with $r=R_1$ in coordinates $(r',\theta',z')$ as shown in panel (b) of Fig. \ref{figcloakto}. Based on this coordinate transformation, the schematic diagram of the ray trajectories in the cloak is shown in panel (c): the rays are distorted and guided around the cloak.

The mapping function is defined as \cite{Pendry06}
\begin{equation}
  \left\{
  \begin{array}{l}
  r'=R_1+r(R_2-R_1)/R_2, \quad 0 \leq r \leq R_2\\
  \theta'=\theta, \quad 0< \theta \leq 2\pi \\
   z'=z\,.
  \end{array}
  \right.
  \label{gtcloak}
\end{equation}
which provides the Jacobian matrix
\begin{equation}
  {\bf J}_{rr'}=\dfrac{\partial (r,\theta,z)}{\partial (r',\theta',z')}={\rm diag} (\dfrac{R_2}{R_2-R_1},1,1) \quad 0 \leq r \leq R_2\,.
\end{equation}

Finally, we go back to the Cartesian coordinates $(x',y',z')$, which are radially contracted, and the compound Jacobian matrix is
\begin{align}
  {\bf J}_{xx'}&={\bf J}_{xr}{\bf J}_{rr'}{\bf J}_{r'x'}
  \label{jm1}
\end{align}
On the other hand, according to the coordinates transformation (\ref{gt1}), we have
\begin{equation}
\begin{array}{l}
  {\bf J}_{xr}=\dfrac{\partial (x,y,z)}{\partial (r,\theta,z)}={\bf R}\,(\theta)\,{\rm diag}(1,r,1)\,\\[2mm]
  {\bf J}_{r'x'}=\dfrac{\partial (r',\theta',z')}{\partial (x,y,z)}={\rm diag}(1,\dfrac{1}{r'},1)\,{\bf R}^{-1}(\theta')
\end{array}
\end{equation}
with
\begin{equation}
  {\bf R}(\theta)=\left[
  \begin{array}{ccc}
  \cos(\theta) & -\sin(\theta) & 0 \\
  \sin(\theta) & \cos(\theta) & 0 \\
  0 & 0 & 1
  \end{array}
  \right]
\end{equation}
Then, (\ref{jm1}) becomes
\begin{align}
  {\bf J}_{xx'}&={\bf J}_{xr}{\bf J}_{rr'}{\bf J}_{r'x'} \notag\\
  &={\bf R}\,(\theta)\,{\rm diag}(1,r,1)\,{\rm diag}\left(\dfrac{R_2}{R_2-R_1},1,1\right)\,{\rm diag}(1,\dfrac{1}{r'},1)\,{\bf R}^{-1}(\theta') \notag \\
  &={\bf R}(\theta)\, {\rm diag} \left(\dfrac{R_2}{R_2-R_1},\dfrac{r}{r'},1\right) \, {\bf R}(-\theta')\,.
\end{align}
The transformation matrix is correspondingly
\begin{equation}
\begin{array}{ll}
  {\bf T}^{-1}_{xx'}&=\left[{\bf J}^{\rm T}_{xx'}{\bf J}_{xx'}/{\rm det}({\bf J}_{xx'})\right]^{-1} \\
  &={\bf R}(\theta')\, {\rm diag}\left(\dfrac{r'-R_1}{r'},\dfrac{r'}{r'-R_1},\dfrac{R_2^2}{(R_2-R_1)^2} \,\dfrac{r'-R_1}{r'}\right) {\bf R}(-\theta')\,.
\end{array}
\label{cloak}
\end{equation}
More explicitly, we apply the matrix ${\bf R}(\theta')$ into (\ref{cloak}) which we denote by
\begin{equation}
  {\bf T}^{-1}_{xx'}= \left[ \begin{array}{ccc}
    T_{11} & T_{12} & 0 \\
    T_{21} & T_{22} & 0 \\
    0 & 0 & T_{33}
  \end{array}
  \right]\,,
  \label{transT}
\end{equation}
with
\begin{equation}
  \begin{array}{l}
    T_{11}=1-\dfrac{R_1}{r'}\cos^2(\theta')+\dfrac{R_1}{r'-R_1}\sin^2(\theta')\,, \\[2mm]
    T_{12}=T_{21}=\dfrac{R_1(R_1-2r')}{r'(r'-R_1)}\sin(\theta')\cos(\theta')\,, \\[2mm]
    T_{22}=1-\dfrac{R_1}{r'}\sin^2(\theta')+\dfrac{R_1}{r'-R_1}\cos^2(\theta') \,,\\[2mm]
    T_{33}=\dfrac{R_2^2}{(R_2-R_1)^2} \dfrac{r'-R_1}{r'} \,.
 \end{array}
 \label{Tcloak}
\end{equation}
Hence, the parameters in the annulus $R_1 < r' \leq R_2$ are
\begin{equation}
  v=v_0 {\bf T}^{-1}_{xx'}\,, \quad v=\ep,\,\mu,\, \xi,\,\zeta \,.
  \label{transpara}
\end{equation}
while the parameters of the outer region $r'> R_2$ are unchanged, and that of the disk $r' \leq R_1$ can have any value without affecting the electromagnetic scattering.

\begin{figure}[!htb]
    \centering
    \includegraphics[width=0.95\textwidth]{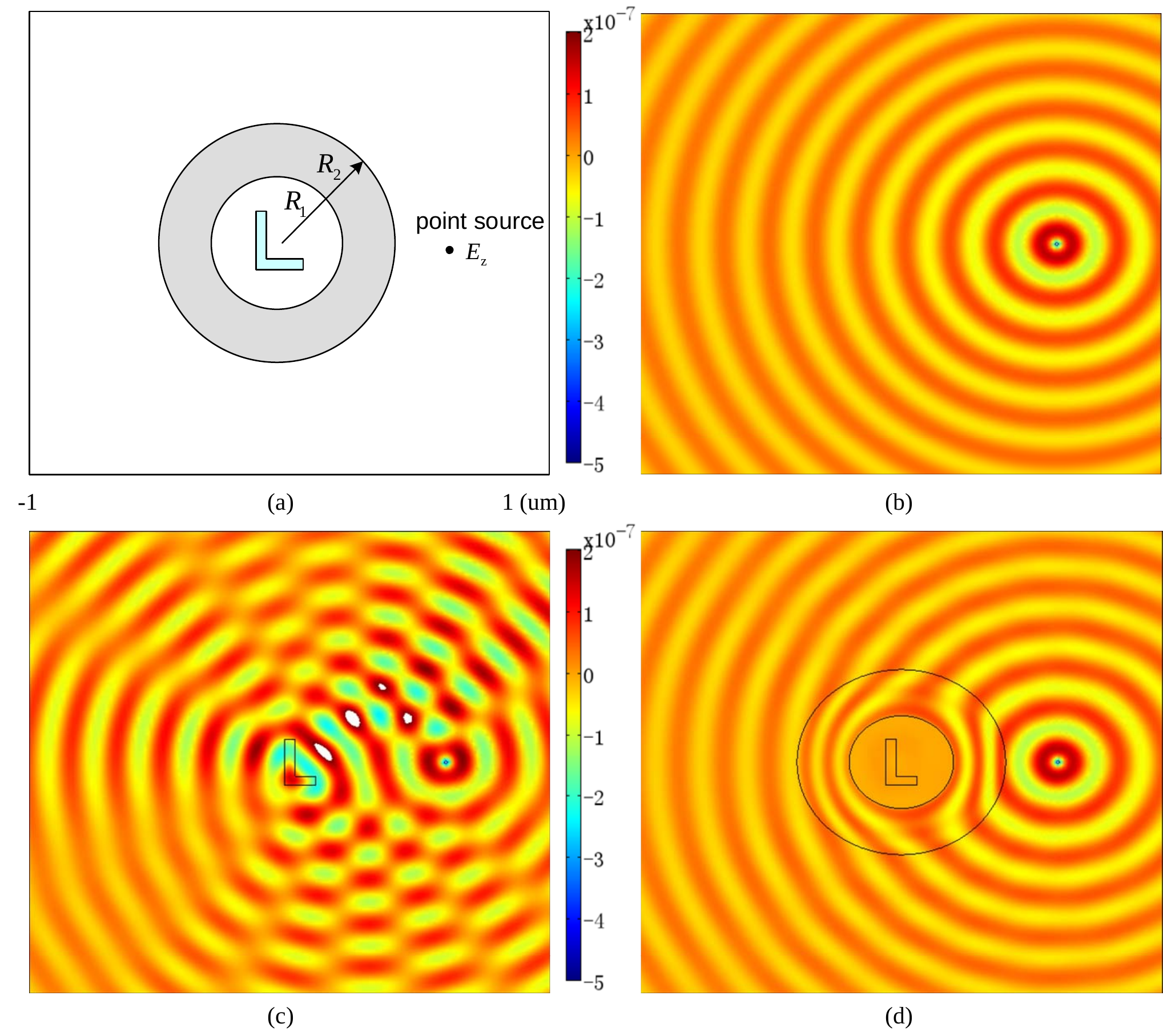}
    \caption{(a) Schematic diagram of bianisotropic cloak with a L-shaped obstacle inside the cloaking region, and a point source ($s$-polarized with the electric field along $z$ with $f=8.7\times10^{14}$Hz) locates outside the shell. Plots of Re$(E_z)$: (b) a point source in a pure matrix with parameters $\ep=\ep_0 {\bf I}_3$, $\mu=\mu_0 {\bf I}_3$ and $\xi=\zeta=0.99/c_0 {\bf I}_3$; (c) same point source in the matrix with a presence of L-shaped obstacle, the parameters of obstacle are $\ep=(1+5\times i)\ep_0{\bf I}_3$, $\mu=\mu_0 {\bf I}_3$ and $\xi=\zeta=0.99/c_0 {\bf I}_3$; (d) same point source that radiates in the presence of an L-shaped obstacle surrounded by an invisibility cloak.}
    \label{figFEMcloak}
\end{figure}

Fig. \ref{figFEMcloak}(a) shows our designed bianisotropic cloak, the cloaking region is the innermost circle of radius $R_1=0.2$um wherein a L-shaped obstacle locates, the grey annulus represents the distorted space with radius $R_1< r' \leq R_2$, where $R_2=0.4$um. Firstly, the EM distribution of a point source ($s$-polarized with the electric field along $z$) with frequency $f=8.7\times10^{14}$Hz radiating in a matrix with parameters $v=v_0 {\bf I}_3,\,(v=\ep,\mu,\xi,\zeta)$ is shown in Fig. \ref{figFEMcloak}(b). $v_0=\ep_0,\mu_0$ are the permittivity, permeability of the vacuum, respectively; while $|\xi_0\zeta_0|\neq 1/c_0$ with $c_0$ the velocity
of light in vacuum should be promised to ensure convergence of the numerical algorithm as indicated in (\ref{condt}). If there is a L-shaped obstacle with $\ep=(1+5\times i)\ep_0{\bf I}_3$, $\mu=\mu_0 {\bf I}_3$, $\xi=\zeta=0.99/c_0 {\bf I}_3$ in the matrix, it leads to an interacting with the source, the plot of Re$(E_z)$ is indicated in panel (c). However, when the L-shaped obstacle is surrounded by the designed cloak, then the obstacle seems to be invisible for the outer observer, as shown in panel (d).

%%---------------------------------------------------------------------------------------------------------------------------------------------------
\subsection{Concentrator}
A cylindrical concentrator is designed to focus the incident waves with wave vectors perpendicular to the cylinder axis, enhancing the electromagnetic energy density in a given area \cite{Rahm08}.

\begin{figure}[!htb]
    \centering
    \includegraphics[width=0.8\textwidth]{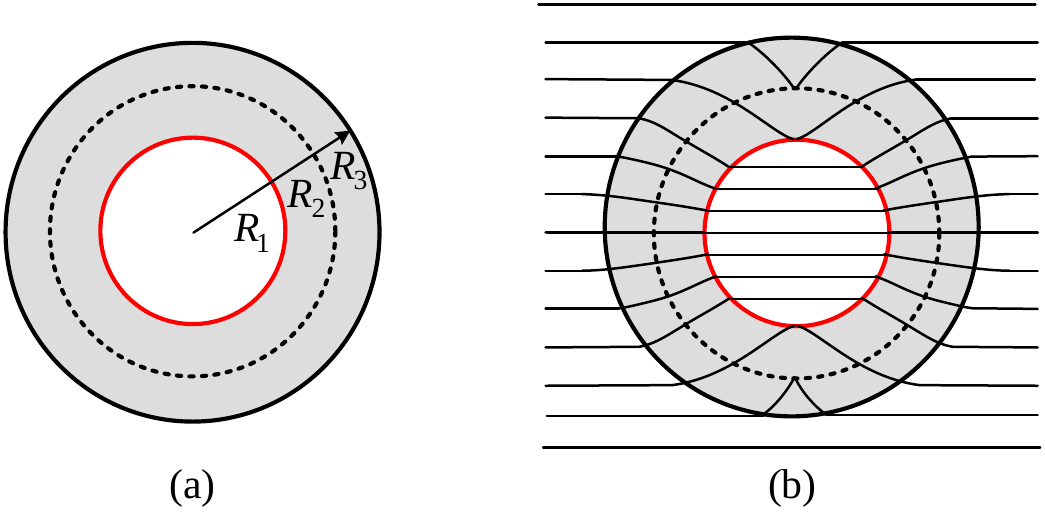}
    \caption{(a) Schematic diagram of a concentrator: a space is compressed into a circle with radius $R_1$ (the innermost in red) with the expense of an expansion of space (grey color) between $R_1$ and $R_3$; the dotted circle corresponds to a virtual interface used in the geometric transformation, see (\ref{rconc}). (b) Ray trajectories in the concentrator.}
    \label{figcr}
\end{figure}

We start with a map from Cartesian coordinates to cylindrical ones, a cylindrical lens with $r=R_3$ is shown in Fig. \ref{figcr}(a), then a radial transformation is introduced to distort the space as
\begin{equation}
\begin{array}{l}
  r'=\left\{ \begin{array}{ll}
    \dfrac{R_1}{R_2}r & 0 \leq r \leq R_2 \\[2mm]
    \dfrac{R_3-R_1}{R_3-R_2}r -\dfrac{R_2-R_1}{R_3-R_2} R_3 & R_2 < r \leq R_3
  \end{array} \right. \\[3mm]
  \theta'=\theta,\quad 0 \leq \theta < 2\pi \\
  z'=z\,.
\end{array}
\label{rconc}
\end{equation}

As shown in Fig. \ref{figcr}(a), a space is compressed into a cylindrical region with radius $R_1$ (the innermost circle) with the expense of an expansion of space(grey color) between $R_1$ and $R_3$, where an intermediate circle is located at $R_2$. Fig. \ref{figcr}(b) shows the ray trajectories in such a concentrator. The associated Jacobian matrix for the transformation in (\ref{rconc}) is
\begin{equation}
  {\bf J}_{rr'}= \left\{ \begin{array}{ll}
    {\rm diag}(\dfrac{R_2}{R_1},1,1) & 0 \leq r \leq R_2 \,,\\[2mm]
    {\rm diag}(\dfrac{R_3-R_2}{R_3-R_1},1,1) & R_2 < r \leq R_3\,.
  \end{array} \right.
\end{equation}
Finally we go back to Cartesian coordinates. The compound Jacobian matrix is then
\begin{equation}
\begin{array}{l}
  {\bf J}_{xx'}={\bf J}_{xr}{\bf J}_{rr'}{\bf J}_{r'x'} \\[2mm]
  =\left\{\begin{array}{ll}
  {\bf R} (\theta') {\rm diag}(\dfrac{R_2}{R_1},\dfrac{r}{r'},1){\bf R} (-\theta')& 0 \leq r \leq R_2 \\[2mm]
  {\bf R} (\theta') {\rm diag}(\dfrac{R_3-R_2}{R_3-R_1},\dfrac{r}{r'},1){\bf R} (-\theta') & R_2 < r \leq R_3
  \end{array} \right.
\end{array}
\end{equation}
Furthermore, the transformation matrix is
\begin{equation}
\begin{array}{l}
  {\bf T}^{-1}_{xx'}={\bf J}^{-1}_{xx'} {\bf J}^{-{\rm T}}_{xx'} \det({\bf J}_{xx'})\\[2mm]
  =\left\{\begin{array}{ll}
  {\bf R} (\theta') {\rm diag}(1,1,\dfrac{R^2_2}{R^2_1}){\bf R} (-\theta')& 0 \leq r \leq R_2 \\[2mm]
  {\bf R} (\theta') {\rm diag}(\dfrac{R_3-R_1}{R_3-R_2}\dfrac{r}{r'},\dfrac{R_3-R_2}{R_3-R_1}\dfrac{r'}{r},\dfrac{R_3-R_2}{R_3-R_1}\dfrac{r}{r'}){\bf R} (-\theta') & R_2 < r \leq R_3
  \end{array} \right.
\end{array}
\end{equation}
Let us expand this formula in the distorted cylindrical coordinates $(r',\theta',z')$, and denote ${\bf T}^{-1}_{xx'}$ as in (\ref{transT}), finally we have
\begin{equation}
  \begin{array}{ll}
    0 \leq r' \leq R_1:& \quad T_{11}=T_{22}=1, \quad
    T_{12}=T_{21}=0,\quad T_{33}=\dfrac{R^2_2}{R^2_1} \,.\\
    R_1 < r' \leq R_3: &\quad T_{11}=A\cos^2(\theta')+B \sin^2(\theta'),\\
    & \quad T_{12}=T_{21}=(A-B)\sin(\theta')\cos(\theta'),\\
    & \quad T_{22}=A\sin^2(\theta')+B \cos^2(\theta')\,, \quad T_{33}=C\,.
  \end{array}
   \label{Tcon}
\end{equation}
with
\begin{equation}
\begin{array}{c}
  A=1+\dfrac{R_2-R_1}{R_3-R_2}\dfrac{R_3}{r'},\quad B=\dfrac{r'}{r'+\dfrac{R_2-R_1}{R_3-R_2}R_3},\\[0.7cm] C=(\dfrac{R_3-R_2}{R_3-R_1})^2(1+\dfrac{R_2-R_1}{R_3-R_2}\dfrac{R_3}{r'})\,.
\end{array}
\end{equation}
Apply matrix ${\bf T}^{-1}_{xx'}$ to (\ref{transpara}), we can obtain the formulae of these parameters in the annulus as well as the innermost circle, while the outside space of the concentrator is bianisotropic matrix with $v=v_0{\bf I}_3$.

\begin{figure}[!htb]
    \centering
    \includegraphics[width=0.95\textwidth]{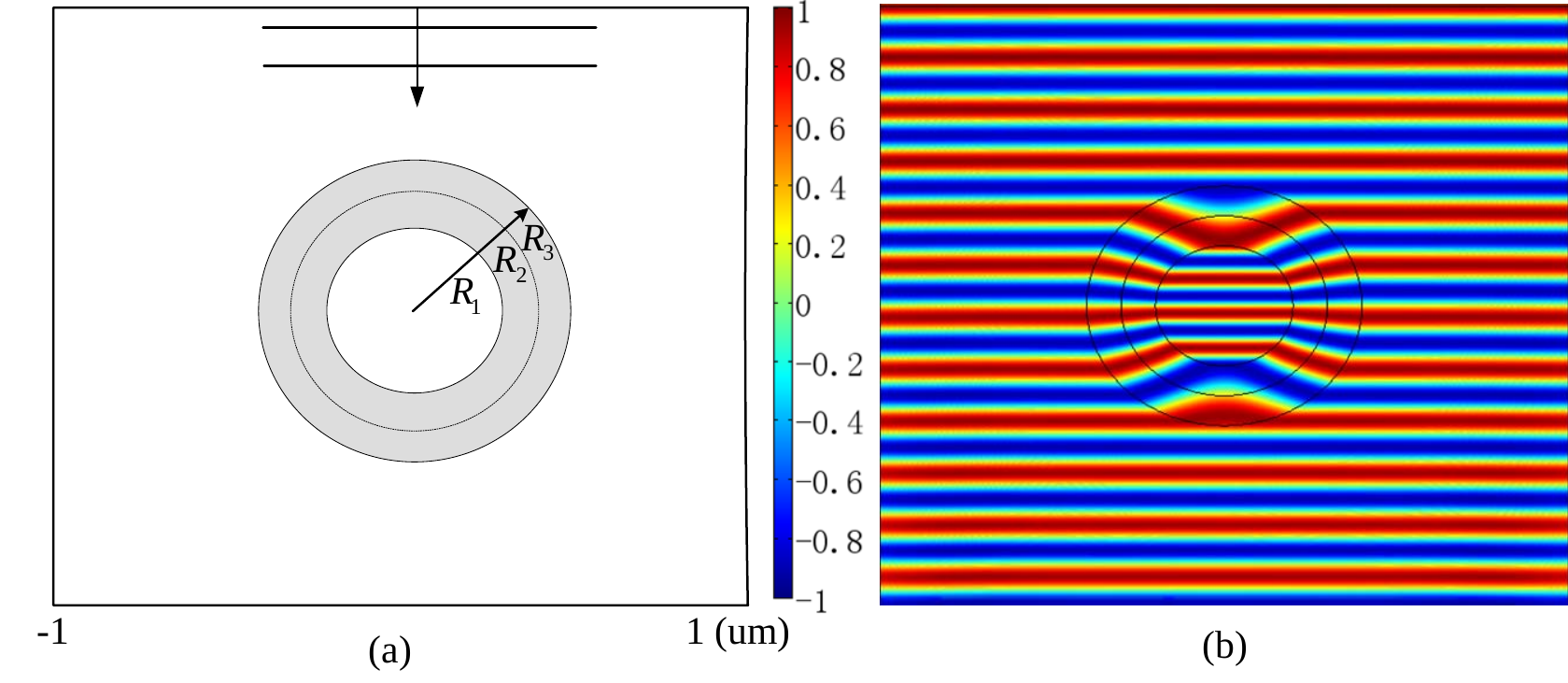}
    \caption{(a) Schematic diagram of a bianisotropic concentrator. (b) Plot of Re$(E_z)$ under an $s$-polarized incidence with frequency $f=8.7\times 10^{14}$Hz, the parameters in the region $r' \leq R_1$, and the annulus $R_1 < r' \leq R_3$ are given by (\ref{transpara}) with transformation matrix ${\bf T}^{-1}_{\bf xx'}$ in (\ref{Tcon}).}
    \label{figFEMcon}
\end{figure}

Fig. \ref{figFEMcon}(a) is the schematic diagram of our concentrator, the radii of the inner and outer circles are $R_1=0.2$um, $R_3=0.4$um, while the intermediate circle has a radius $R_2=0.3$um. The parameters of each region are defined in (\ref{transpara}) along with transfer matrix ${\bf T}^{-1}_{\bf xx'}$ in (\ref{Tcon}), here the magnetoelectric coupling parameters $\xi_0=\zeta_0=0.99/c_0$ are chosen as the same values as those of the invisibility cloak. An $s$-polarized (the electric field is perpendicular to the $x$-$y$ plane) plane wave with frequency $8.7\times 10^{14}$Hz is incident from above, the plot of Re$(E_z)$ is depicted in panel (b): the waves are compressed in the inner circle as expected.
%%--------------------------------------------------------------------------------------------------------------------------------------------------

\subsection{Rotator}
\begin{figure}[!htb]
    \centering
    \includegraphics[width=0.8\textwidth]{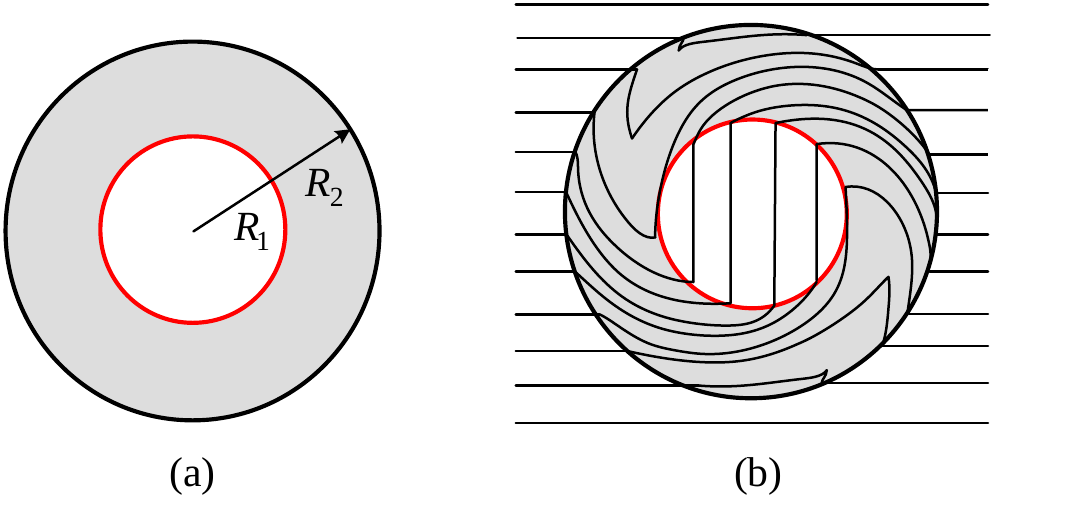}
    \caption{(a) Schematic diagram of the rotator: incident waves are rotated in the annulus $R_1 < r \leq R_2$ and reach the disk of radius $R_1$ with a $\pi/2$ degree rotation. (b) Ray trajectories in rotator. }
    \label{figrot}
\end{figure}

Fig. \ref{figrot} shows a rotator allowing a rotation of the incident wave, the distribution of the field inside a disk $r\leq R_1$ appears as it comes from a different angle comparing with that of outside $r > R_2$, for example, a $\pi/2$ angle rotation is achieved here in the inner disk.

Be different from the geometric transformation introduced in the concentrator, an angular transformation is introduced in this case instead of the radial transformation. The mapping is performed in the region $r < R_2$ : the domain in $r < R_1$ is transformed by rotating a fixed angle from the region $r'=a$ in virtual space, while the rotation angle is continually changed from the fixed angle to zero in region $R_1 < r < R_2$. In other words, the transformation algorithm can be expressed as
\begin{equation}
  \begin{array}{ll}
    r \leq R_1: & r'=r,\quad \theta'=\theta+\theta_0, \quad z'=z \,.\\
    R_1 < r \leq R_2: & r'=r,\quad \theta'=\theta+\theta_0 \dfrac{f(R_2)-f(r)}{f(R_2)-f(R_1)}, \quad z'=z \,.\\
    r > R_2: & r'=r,\quad \theta'=\theta,\quad z'=z\,. \\
  \end{array}
\end{equation}
where $\theta_0$ is the rotation angle for the inner disk, and it is reduced to zero as the radius approaches to $r=R_2$. $f(r)$ is an arbitrary continuous function of $r$.

Based on this transformation, we can derive the associated Jacobian matrix as
\begin{equation}
  {\bf J}_{rr'}= \left\{ \begin{array}{ll}
    {\rm diag}(1,1,1) & r \leq R_1 \quad {\rm and} \quad r > R_2\\[2mm]
   \left[ \begin{array}{ccc}
    1 & 0 & 0 \\
    \theta_0 \dfrac{f'(r)}{f(R_2)-f(R_1)} & 1 & 0\\
    0 & 0 & 1
    \end{array}\right] & R_1 < r \leq R_2
  \end{array} \right.
\end{equation}
again, the compound Jacobian matrix can be obtained by ${\bf J}_{xx'}={\bf J}_{xr}{\bf J}_{rr'}{\bf J}_{r'x'}$, as well as the transformation matrix ${\bf T}^{-1}_{xx'}={\bf J}^{-1}_{xx'} {\bf J}^{-{\rm T}}_{xx'} \det({\bf J}_{xx'})$. In annulus $R_1 < r'(=r) \leq R_2$, the notations in (\ref{transT}) become
\begin{equation}
  \begin{array}{l}
    T_{11}=1+r'^2 D^2 \sin^2(\theta')+2r'D \sin(\theta')\cos(\theta') ,\\[2mm]
    T_{12}=T_{21}=r'D[\sin^2(\theta')-\cos^2(\theta')]-r'^2D^2 \sin(\theta')\cos(\theta')\,, \\[2mm]
    T_{22}=1+r'^2 D^2 \cos^2(\theta')-2r'D \sin(\theta')\cos(\theta'),\quad T_{33}=1\,.
  \end{array}
   \label{Trot}
\end{equation}
with $D =\theta_0 \dfrac{f'(r)}{f(R_2)-f(R_1)}$. In regions $r'\leq R_1$ and $r' \geq R_2$,  ${\bf T}^{-1}_{xx'}={\bf I}_3$. Again, substituting the matrix ${\bf T}^{-1}_{xx'}$ into (\ref{transpara}), the expressions for those parameters can be derived in different regions.

\begin{figure}[!htb]
    \centering
    \includegraphics[width=0.95\textwidth]{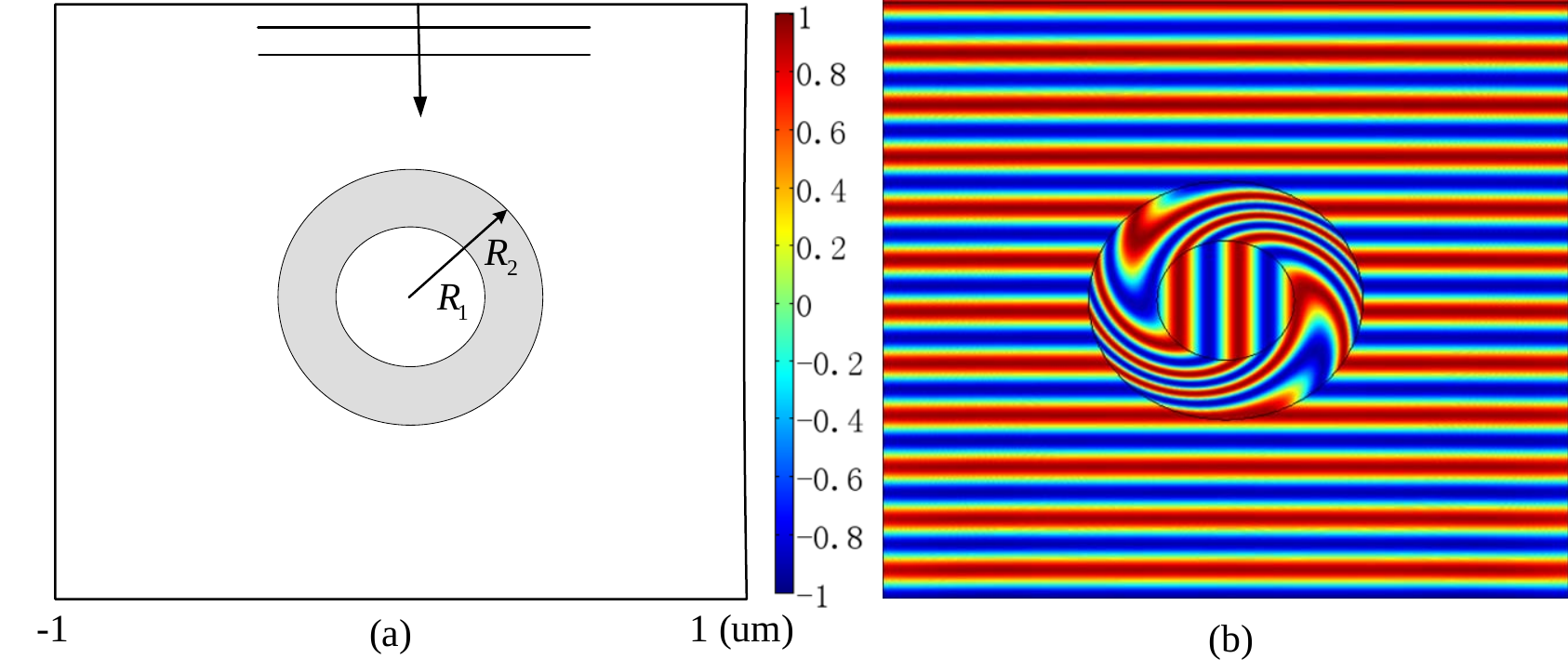}
    \caption{(a) Schematic diagram of the bianisotropic rotator. (b) Plot of Re$(E_z)$ under an $s$-polarized incidence with frequency $f=8.7\times 10^{14}$Hz, the parameters in the region $r' \leq R_1$, annulus $R_1 < r' \leq R_2$ are given by (\ref{transpara}) with transformation matrix ${\bf T}^{-1}_{xx'}$ in (\ref{Trot}).}
    \label{figFEMrot}
\end{figure}
Similarly, the FEM analysis for the rotator is shown in Fig. \ref{figFEMrot}.  The radii of the inner and outer circles are $R_1=0.2$um, $R_2=0.4$um, respectively. The parameters of each region are defined in (\ref{transpara}) along with transfer matrix ${\bf T}^{-1}_{xx'}$ in (\ref{Trot}). An $s$-polarized incident wave with frequency $f=8.7\times10^{14}$Hz is assumed to radiate from above. A rotation of $\theta_0=\pi/2$ of the fields in the inner circle can be observed in panel (b) comparing with the incidence. Note that, $\theta_0$ can be an arbitrary values in $[-\pi,\pi]$.
%---------------------------------------------------------------------------------------------------------------------------

\section{A multilayered bianisotropic cloak through homogenization}
Let us now consider a layered medium periodic along $x_1$ and invariant along $x_2$ and $x_3$. A periodic cell consists of a heterogeneous bi-isotropic medium described by piecewise constant scalar valued functions $\varepsilon(x_1)$, $\mu(x_1)$, $\xi(x_1)$, $\zeta(x_1)$. Using two-scale expansion techniques applied to the Maxwell-Tellegen's equations like in \cite{seb2007} one gets the homogenized equations
\begin{equation}
\begin{array}{lll}
\nabla \times {\bm E}_{\rm eff} &= \omega \zeta_{\rm eff} {\bm E}_{\rm eff}+i \omega \mu_{\rm eff} {\bm H}_{\rm eff}\,,\\
\nabla \times {\bm H}_{\rm eff} &= -i \omega \varepsilon_{\rm eff} {\bm E}_{\rm eff}+\omega \xi_{\rm eff}{\bm H}_{\rm eff} \,.\\
\end{array}
\label{f5a}
\end{equation}
where
$\varepsilon_{\rm eff}$, $\mu_{\rm eff}$, $\xi_{\rm eff}$ and $\zeta_{\rm eff}$ are anisotropic
homogenized tensors such that
\begin{equation}
\begin{array}{l}
\varepsilon_{\rm eff} =\left[\begin{array}{ccc}
{<\varepsilon^{-1}>}^{-1} & 0 & 0\\
0 & <\varepsilon> & 0\\
0 & 0 & <\varepsilon>
\end{array}\right] , \\[1cm]
\mu_{\rm eff} =\left[\begin{array}{ccc}
{<\mu^{-1}>}^{-1} & 0 & 0\\
0 & <\mu> & 0\\
0 & 0 & <\mu>
\end{array}\right] , \\[1cm]
\xi_{\rm eff} =\left[\begin{array}{ccc}
{<\xi^{-1}>}^{-1} & 0 & 0\\
0 & <\xi> & 0\\
0 & 0 & <\xi>
\end{array}\right]\,, \\[1cm]
\zeta_{\rm eff} =\left[\begin{array}{ccc}
{<\zeta^{-1}>}^{-1} & 0 & 0\\
0 & <\zeta> & 0\\
0 & 0 & <\zeta>
\end{array}\right]\,.
\end{array}
\label{f10}
\end{equation}
where $<f(x_1)>=\int_0^d f(x_1) \, dx_1$, with $d$ the periodic cell size.
Note also that these formulae can be deduced from the high-order homogenization approach in \cite{yanPRSA13}, by keeping only leading
order terms.

One can use these formulae to design a concentric multilayered bianisotropic cloak of inner radius $R_1=140$nm and outer
radius $R_2=300$nm, consisting of 22 homogeneous layers (see table \ref{tab1} for the values of the inverse relative permittivity $\varepsilon_r$, permeability $\mu_r$, magnetoelectric coupling parameters $\xi_r$ and $\zeta_r$) in a bianisotropic background with permittivity $\varepsilon=\varepsilon_0$, permeability $\mu=\mu_0$ and magnetoelectric coupling parameters $\xi=\xi_0=0.99/c_0$, $\zeta=\zeta_0=0.99/c_0$.
\begin{table}[h]
\begin{center}
\caption{Parameters of the layered cloak.\,($\varepsilon=\varepsilon_r \varepsilon_0, \mu=\mu_r \mu_0,\xi=\xi_r\xi_0,\zeta=\zeta_r\zeta_0$)}
\begin{tabular}{|l|c|c|c|c|c|c|}
\hline
layer & 1 (inner)& 2 & 3 &4 & 5 & 6   \\
\hline
thickness [nm]  & 10.0 & 3.75 & 7.5 & 7.5 &
7.5 & 7.5 \\
\hline
$\varepsilon^{-1}_r = \mu^{-1}_r = \zeta^{-1}_r = \xi^{-1}_r$ & 2.0 & 1680.7 & 0.2503 & 80.7492 &
0.2514 & 29.3898 \\
\hline
layer & 7& 8& 9 & 10 & 11 & 12  \\
\hline
thickness [nm]  & 7.5 & 8.75 & 6.25 & 7.5 &
7.5 & 7.5 \\
\hline
$\varepsilon^{-1}_r = \mu^{-1}_r = \zeta^{-1}_r = \xi^{-1}_r $ & 0.2530 & 16.3674 & 0.2548 & 10.9864 & 0.2568 &
8.1798 \\
\hline
layer & 13&14&15&16& 17 & 18   \\
\hline
thickness [nm]  & 7.5 & 7.5 & 7.5 & 7.5 &
7.5 & 7.5 \\
\hline
$\varepsilon^{-1}_r = \mu^{-1}_r = \zeta^{-1}_r = \xi^{-1}_r $ & 0.2589 & 6.5 & 0.2611 & 5.3990 & 0.2632 & 4.6292 \\
\hline
layer & 19 & 20 & 21 & 22 &  &   \\
\hline
thickness [nm]  & 7.5 & 7.5 & 7.5 & 3.75 &  &  \\
\hline
$\varepsilon^{-1}_r = \mu^{-1}_r = \zeta^{-1}_r = \xi^{-1}_r $  & 0.2653 & 4.0645 & 0.2674 & 1.0 & & \\
\hline
\end{tabular}
\label{tab1}
\end{center}
\end{table}

Notice that almost all layers have an equal thickness of 7.5 nm and one layer in two has almost the same optical
properties.

\begin{figure}[!htb]
    \centering
    \includegraphics[width=0.95\textwidth]{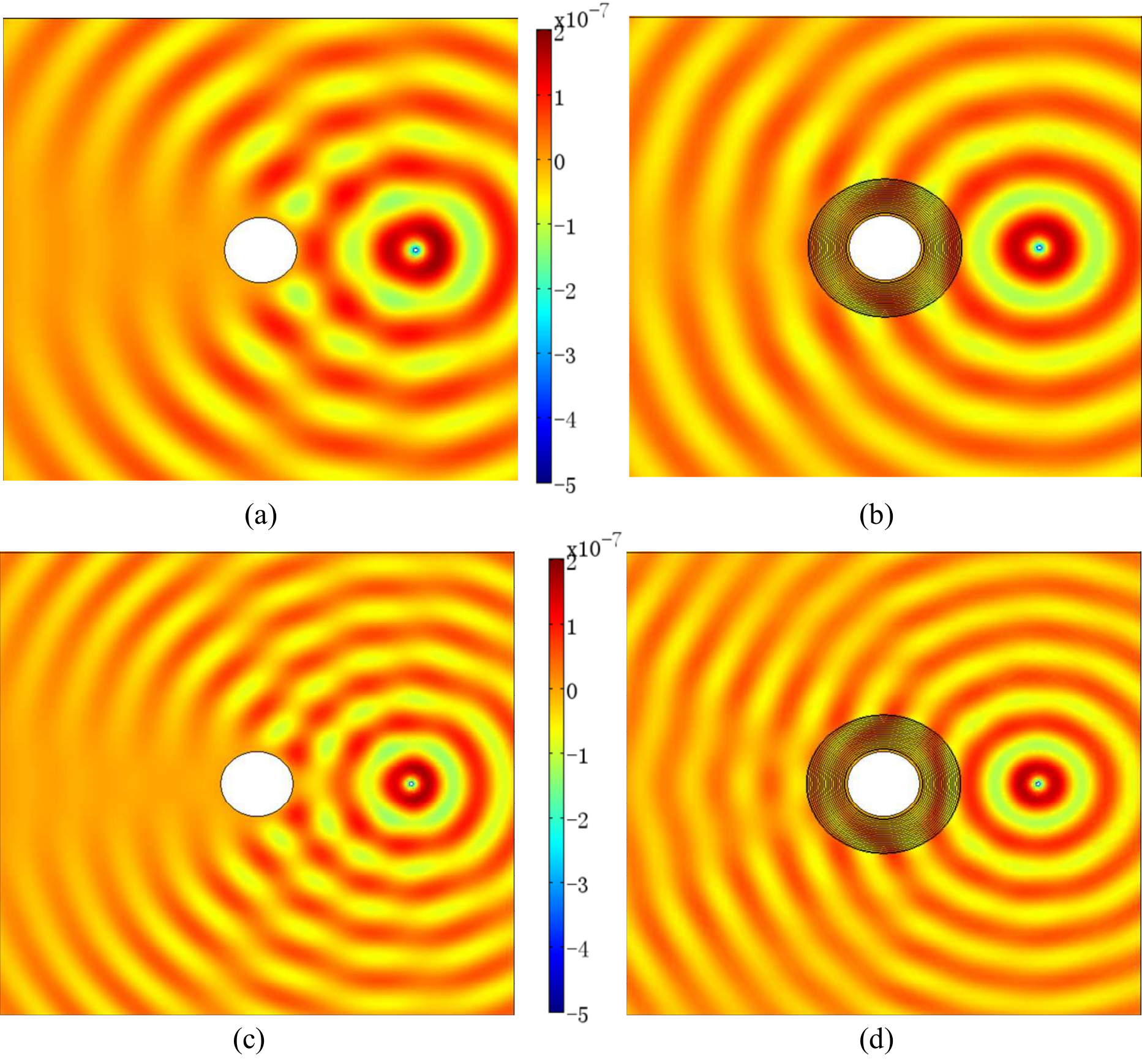}
    \caption{(a) Scattering of an infinite conducting obstacle (of radius $140$ nm) by a point source at a frequency $570$ THz; (b) Scattering like in (a) when the obstacle is surrounded by a layered cloak. (c) Scattering like in (a) for a point source at a frequency $770$ THz; (d) Scattering like in (b) for a point source at a frequency $770$ THz. Color scale corresponds to normalized values of Re$(E_z)$.}
    \label{figlast}
\end{figure}

We show the result of numerical simulations in Fig. \ref{figlast}.One can see that the backscattering by the infinite conducting
obstacle (of radius $140$ nm) with a concentric multilayered bianisotropic cloak is much reduced in Fig. \ref{figlast}(b) comparing with Fig. \ref{figlast}(a), and the shadow zone behind the obstacle is also much reduced with the layered cloak. The interval of cloaking frequencies from $570$ to $770$ THz is fairly broadband.

\section{Conclusion}
In this paper, we have discussed the PDEs model for the analysis of the propagation properties in bianisotropic media with an invariance along one direction: two coupled PDEs with scalar solutions are derived for this electromagnetic system. The spatially varying coefficients within these two PDEs can be scalar or matrix valued. Regarding the numerical model, specially designed perfectly matched layers (PMLs) are introduced in order to account for the unbounded domain. To do this, we consider a transformation which maps an infinite domain onto a finite one, and we further add the damping. This provides a reflection-less bianisotropic layer for all incident angles, and this can be considered as an extension of the PMLs introduced by J. Berenger nearly twenty years ago \cite{Berenger94}. We implement these PMLs in scattering problems in bianisotropic media including an invisibility cloak, a field concentrator and a field rotator designed by introducing the proper geometric transformation. Importantly, these functional structures are proved to work well when they are achieved by bianisotropic media. We finally, show that one can achieve near-cloaking with a layered cloak with piecewise constant permittivity, permeability and magnetoelectric coupling parameters. The solutions of singular systems considered in this paper represent important benchmarks for the validation of non-trivial numerical calculations for bianisotropic media.

\section{Acknowledgments}
S.Guenneau acknowledges a European funding through ERC Starting Grant No. 279673.

\section*{References}

\bibliography{mybibfile}

\end{document}